\documentclass[12pt,a4paper]{article}
\usepackage{amsfonts,amsmath}
\usepackage[mathscr]{eucal}
\usepackage{hyperref}

\def\a {\alpha}

\def\l {\lambda}

\def  \Ber {{\rm Ber\,}}
\def  \Tr {{\rm Tr\,}}
\def \ZZ{{\mathbb{Z}}}

\title  {New Facts about Berezinians\thanks{Based on a talk given at the International Workshop
``Supersymmetries and Quantum Symmetries", Dubna, July 27--31, 2005}}
\newcommand{\myaddress}{{\footnotesize
$^{\scriptstyle{1}}$ School of Mathematics, University of
Manchester, Sackville Street,\\
Manchester, M60 1QD, UK\\ \smallskip} {\footnotesize
$^{\scriptstyle{2}}$ G.~S.~Sahakian~Department~of~Theoretical
~Physics, Yerevan State University, A.~Manoukian Street, 375049 Yerevan, Armenia\\ \smallskip}
{\footnotesize {\tt
khudian@manchester.ac.uk, theodore.voronov@manchester.ac.uk} }}

\author{Hovhannes M. Khudaverdian{\small $^{\scriptstyle{1,2}}$},
Theodore  Th. Voronov{\small $^{\scriptstyle{\,1}}$}}

\date{\myaddress}

\begin {document}
\maketitle
\begin{abstract} We consider a new formula for Berezinian (superdeterminant). The Berezinian
of a supermatrix $A$ is expressed as the ratio of polynomial
invariants of $A$. This formula follows from recurrence relations
existing for supertraces of exterior powers.
\end{abstract}

 Tools of supermathematics have become an essential part of the
   mathematical baggage of theoretical physics.
 On the other hand, still there are many important  questions corresponding
 to  statements well-known in the ordinary case, answers to which are
 not at all clear in the supercase.

  Here we discuss some of these problems.
   We consider deep relations that arise in the supercase between Berezinian (superdeterminant)
   and exterior powers. In particular, this allows to give a new expression for the Berezinian
   in terms of polynomial invariants of a matrix. Details see in our paper \cite{khudvoron:berez}.

  Recall the relations between traces and (ordinary) determinant.
  If $A$ is a $2\times 2$ matrix   $\begin{pmatrix}
    a & b \\ c & d \\
  \end{pmatrix}$,
$\det
A=ad-bc=\frac{1}{2}\left(\left(a+d\right)^2-\left(a^2+2bc+d^2\right)\right)$
                         $=
\frac{1}{2}\left(\Tr^2 A-\Tr A^2\right)
    $
  and
             $
  \det (1+Az)=1+z\Tr A+z^2\det A.
             $
 This is a commonplace. In general, if $A$ is an $n\times n$ matrix,
 then one can consider the following  polynomial of degree $n$:
 \begin{equation}\label{generpolynomial}
   R_A(z)=\det (1+Az)=\sum_{k=0}^n c_k(A)z^k\,,
 \end{equation}
 the characteristic polynomial of the matrix $A$.
   (For our purposes it is more convenient to consider the above polynomial  instead
  of  $\det (A-z)$.)
  One can easily  calculate the coefficients of this polynomial by taking the derivative  with respect to
  $z$.
 We arrive at the relation:
\begin{multline*}
    {d\over dz}\det (1+Az)=\sum_{k=1}^n kc_k(A)z^{k-1}= \\
    \det (1+Az){\Tr \left((1+Az)^{-1}A\right)}=
   \sum_{k=0}^n c_k(A)z^k\,\sum_{k=0}^{\infty} (-1)^ks_{k+1}z^k  \,,
\end{multline*}
 where we denoted $s_k(A)=\Tr A^k$. This leads to recurrence relations expressing $c_k(A)$
 in terms of $s_k(A)$:
\begin{equation}\label{recurrel1}
c_0=1,\, c_1=s_1,\   \dots,\  c_{k+1}=\frac{1}{k+1}(s_1 c_k- s_2
c_{k-1}+\ldots +(-1)^ks_{k+1}),\  \ldots\
   \end{equation}
In particular,
    \begin{equation}\label{determinantviatrace}
    \det A=c_n(A)\,\, \text{\  if $A$ is an operator on an $n$-dimensional space,}
 \end{equation}
 and it can be expressed via $s_k(A)$.
 \noindent These are standard facts in linear algebra.

  What about a generalisation of the above formulae  to the supercase?

 Let $V$ be a $p|q$-dimensional superspace.
 One can describe it in the following way.  Let
 $V_0\oplus V_1$ be the direct sum of $p$-dimensional and $q$-dimensional vector spaces.
 Let $\{{\bf e}_i\}$
 ($i=1,\dots,p$) and $\{{\bf f}_\alpha\}$ ($\a=1,\dots,q$) be bases in the spaces
 $V_0, V_1$, respectively.
 Consider linear combinations $\sum_{i=1}^p a^i{\bf e}_i+\sum_{\alpha=1}^q b^\alpha{\bf f}_\alpha$
 where coefficients $a^i$ are even elements of some  Grassmann algebra $\Lambda$
 and $b^\alpha$ are odd elements of this  Grassmann algebra.
  Such linear combinations are considered as points of the $p|q$-dimensional superspace $V$.

 Let  $A$ be an even linear operator on this space.
 The (super)matrix of the operator $A$ has the form $\begin{pmatrix}
                                                       A_{00} & A_{01} \\
                                                        A_{10} & A_{11} \\
                                                     \end{pmatrix}$,
where $A_{00}, A_{11}$ are $p\times p$ and $q\times q$ matrices,
respectively,  with even entries
   taken from the Grassmann algebra $\Lambda$, and
   $A_{01}, A_{10}$ are $p\times q$ and $q\times p$ matrices, respectively, with odd entries
   from the Grassmann algebra $\Lambda$.  Such a (super)matrix is called even.

    The Berezinian (superdeterminant) of an even matrix $A$
   is given by the famous formula  due to F.~A.~Berezin (see \cite{berezin:antire}):
                \begin{equation}\label{berezinian}
    \Ber A=\frac{\det \left (A_{00}-A_{01}A_{11}^{-1}A_{10}\right)}{\det A_{11}}\,.
\end{equation}
Berezinian is a multiplicative function of matrices, $\Ber (AB)=\Ber
A\cdot \Ber B$. Hence $\Ber$ is well-defined on operators.
Berezinian is related with supertrace in the same way as the
ordinary determinant, with  trace: for an even supermatrix  $D$
\begin{equation}\label{bertracerelations}
    \Ber e^D=e^{\Tr D}\,.
\end{equation}
We denote the supertrace of a supermatrix by the same symbol as the trace of an ordinary matrix. Recall
that for an even supermatrix
                    $$
                    \Tr D=
         \Tr \begin{pmatrix}
               D_{00} & D_{01} \\
               D_{10} & D_{11} \\
             \end{pmatrix}=
                                 \Tr D_{00}-\Tr D_{11}\,.
                    $$

 Instead of the characteristic polynomial \eqref{generpolynomial} one has to consider the
 {\it characteristic rational function}
 $R_A(z)=\Ber (1+Az)$. We note that the  straightforward use of  expression \eqref{berezinian}
 for the analysis of the characteristic function leads to a confusion.

 Let us step back and consider the geometrical meaning of coefficients $c_k(A)$
 in  formula \eqref{generpolynomial} for the ordinary case. Suppose that
   $\{{\bf e}_i\}$ is an eigenbasis of a linear operator $A$ on an $n$-dimensional space $V$:
$A{\bf e}_i=\l_i {\bf e}_i$ ($i=1,\dots,n$).
 Then \begin{equation*}
R_A(z)=\det (1+Az)=\prod_{i=1}^n(1+\l_iz)=\sum_{k=0}^n
\left(\prod_{j_1<j_2<\dots<j_k}\l_{j_1}\dots\l_{j_k}\right)z^k=
\sum_{k=0}^n c_k(A)z^k\,.
\end{equation*}
Consider the basis consisting of wedge products $\{{\bf
e}_{j_1}\wedge \dots \wedge{\bf e}_{j_k}\}$ ($1\leq
j_1<j_2<\dots<j_k\leq n$) in the exterior power $\wedge^k V$.
   Then $\l_{j_1}\dots\l_{j_k}$ is the eigenvalue corresponding to the basis vector
${\bf e}_{j_1}\wedge \dots \wedge{\bf e}_{j_k}$. Hence we see that
for the polynomial $\det (1+Az)$,
\begin{equation*}\label{classicham1}
        c_k(A)=\Tr \wedge^k A,\,\,
           \end{equation*}
   where we denote by $\wedge^k A$ the operator induced
by $A$ in the exterior power $\wedge^k V$.

 This formula can be straightforwardly generalised
   to the supercase (see \cite{schmitt:ident}, \cite{khudvoron:berez}).
Suppose that   $\{{\bf e}_i,{\bf f}_\a\}$ is an eigenbasis of a
 linear operator $A$ in a $p|q$-dimensional superspace $V$:
$A{\bf e}_i=\l_i {\bf e}_i, A{\bf f}_\a=\mu_\a {\bf f}_\a$.
Here $\{{\bf e}_i\}$, $i=1,\dots,p$, are even eigenvectors and
$\{{\bf f}_\a\}$, $\a=1,\dots,q$, are odd eigenvectors.
 Then
    \begin{multline}\label{generpolynomialdiag}
R_A(z)=\Ber (1+Az)=\sum_{k=0}^{\infty}c_k(A)z^k=
\prod_{i=1,\a=1}^{i=p,\a=q} \frac{1+\l_i z}{1+\mu_\a z}=\\
\sum_{r=0}^p\sum_{s=0}^\infty
\left(\prod_{j_1<j_2<\dots<j_r}\l_{j_1}\dots\l_{j_r}\right)z^r
\left(\prod_{\beta_1\leq
\beta_2\leq\dots\leq\beta_s}(-1)^s\mu_{\beta_1}\mu_{\beta_2}\dots\mu_{\beta_s}\right)z^s\,.
      \end{multline}
Consider the basis $\{{\bf e}_{j_1}\wedge \dots \wedge {\bf
e}_{j_r}\wedge {\bf f}_{\beta_1}\wedge\dots\wedge{\bf
f}_{\beta_s}\}$ ($1\leq j_1<j_2<\dots<j_k\leq
p,1\leq\beta_1\leq\dots\leq\beta_s\leq q$, $r+s=k$) in the exterior
power $\wedge^k V$.
 Then
$\l_{j_1}\dots\l_{j_r}\mu_{\beta_1}\dots\mu_{\beta_s}$ is the
eigenvalue corresponding to  the basis vector ${\bf e}_{j_1}\wedge
\dots \wedge {\bf e}_{j_r}\wedge {\bf
f}_{\beta_1}\wedge\dots\wedge{\bf f}_{\beta_s}$. Hence in the same
way as above the coefficients $c_k(A)$ of the expansion of the
characteristic function at zero give traces of the exterior powers:
              \begin{equation}\label{superham1}
 R_A(z)=\Ber(1+Az)=\sum_{k=0}^{\infty}c_k(A)z^k,\,\, \text{where} \,\,
c_k(A)=\Tr \wedge^k A\,\,  (k=0,1,2,\dots)\,.
           \end{equation}
  Relations \eqref{recurrel1} between $c_k(A)$ and $s_k(A)=\Tr A^k$ remain the same as in
 the ordinary case  because of \eqref{bertracerelations}.
 The essential difference is that now $R_A(z)$ is a fraction, not a
 polynomial as in~\eqref{generpolynomial};
 there are infinitely many terms $c_k(A)$ in the power expansion \eqref{superham1}.

Consider now the expansion  of the characteristic function $R_A(z)$
at infinity. It  leads to traces
  of the exterior powers of the inverse matrix. Indeed, $\Ber (1+Az)=z^{p-q}\Ber A \cdot\Ber (1+A^{-1}z^{-1})$.
   From  \eqref{superham1} it follows that
            \begin{multline}\label{superhamininfinity}
    R_A(z)=z^{p-q}\Ber A\cdot \Ber (1+A^{-1}z^{-1})=z^{p-q}\Ber A\sum_{k=0}^\infty c_k(A^{-1})z^{-k}=\\
          \sum_{k\leq p-q}\left(\Ber A\cdot c_{p-q-k}(A^{-1})\right)z^k=
          \sum_{k\leq p-q}c_k^*(A)z^k\,
         \end{multline}
         near infinity,
 where we have denoted  by\begin{equation}\label{denoteby}
    c_k^*(A)=\Ber A \cdot c_{p-q-k}(A^{-1})=\Ber A\cdot \Tr\wedge^{p-q-k}A^{-1},\,(k=p-q,\,p-q-1,\dots)\,.
           \end{equation}
 The coefficient
 $c_k^*(A)$  can be interpreted as the trace of the representation on the space $\Ber V\otimes \wedge^{p-q-k}V^*$.

In the ordinary case when $V$ is an $n$-dimensional vector space so
that $p=n,q=0$,  then  both \eqref{superham1} and
\eqref{superhamininfinity} are the same polynomial. Comparing them,
we see that
 \begin{equation}\label{identityforminors}
    c_k(A)=\Tr \wedge^k A=c^*_k(A)=\det A\cdot\Tr  \wedge^{n-k} A^{-1}\,.
\end{equation}
 This is a well-known identity between minors of the matrix $A$ and its inverse $A^{-1}$.
 In particular,  for $k=n$  we arrive at \eqref{determinantviatrace}.
  Relation \eqref{identityforminors} holds for any invertible operator $A$.
 This is due  to a canonical isomorphism existing in the ordinary case
  between the spaces $\wedge^k V$ and $\det V \otimes \wedge^{n-k} V^*\,$:
\begin{equation}\label{canisomorphism}
    \wedge^k V\approx\det V\otimes \wedge^{n-k} V^*\,.
\end{equation}

What happens in the supercase?
 Both expansions~\eqref{superham1} and~\eqref{superhamininfinity} are infinite series.
 Claim: the coefficients of both series form {\bf recurrent sequences}.
 Indeed,  we see from \eqref{generpolynomialdiag}
 that the function $R_A(z)$ is the ratio of two polynomials  of degrees
$p$ and $q$, respectively:
       \begin{equation}\label{ratiooftwopolynomials}
    R_A(z)=\Ber(1+Az)=\frac{P(z)}{Q(z)}=\frac{1+a_1z+a_2z^2+\dots+a_pz^p}{1+b_1z+b_2z^2+\dots+b_qz^q}\,.
                     \end{equation}
  Comparing this fraction with the expansion of $R_A(z)$ around zero
  we arrive at the recurrence relations
   \begin{equation}\label{recurrentrealtionsright}
    c_{k+q}+b_1c_{k+q-1}+\dots+b_qc_k=0
\end{equation}
satisfied for all $k>p-q$. Comparing the fraction
in~\eqref{ratiooftwopolynomials} with the expansion of $R_A(z)$
around infinity
  we again arrive at  recurrence relations:
   \begin{equation*}\label{recurrentrealtionsleft}
    c^*_{k}+b_1c_{k-1}^*+\dots+b_qc_{k-q}^*=0
     \end{equation*}
  satisfied for all $k<0$.
 We see that both sequences $\{c_k(A)\}$ and $\{c_k^*(A)\}$
 satisfy the same recurrence relations of  order $q$. It is convenient
 to consider these sequences for all integer $k$ by setting $c_k=0$ for all $k<0$ and $c^*_k=0$ for all
 $k>p-q$. Combine these two sequences
 in one sequence by considering the differences:
       \begin{equation*}\label{theorem}
    \gamma_k=c_k-c_k^*\,.
\end{equation*}
   The sequence $\{\gamma_k\}$ satisfies the same recurrence relations {\bf for all integer} $k$:
   \begin{equation*}\label{recurrentrealtionsforgamma}
    \gamma_{k}+b_1\gamma_{k-1}+\dots+b_q\gamma_{k-q}=0,\,\text{\quad for all $k$}\,.
   \end{equation*}
  Note that in this formula the terms $c_k=\Tr \wedge^k A$ and $c_k^*=\Ber A\cdot \Tr^{p-q-k} A^{-1}$
  are simultaneously  non-zero only in  a finite range where $k=0,1,\dots,p-q$.
  Otherwise $\gamma_k=c_k-c_k^*$ equals
    either $c_k(A)$ for $k>p-q$ or $-c_k^*$ for $k<0$.

The condition that $\{\gamma_k\}$ is a recurrent sequence of  order $q$
can be rewritten in the following closed form:
\begin{equation}\label{identity}
    \det\begin{pmatrix}
          \gamma_k & \dots & \gamma_{k+q} \\
  \dots & \dots & \dots \\
  \gamma_{k+q} & \dots & \gamma_{k+2q} \\
        \end{pmatrix}=0\,\, \text{\quad for all $k\in\ZZ$}\,.
\end{equation}

   Using relations \eqref{recurrentrealtionsright} for $c_k$ only,  one
 can reconstruct the function $R_A(z)$ and all rational invariants
 of the matrix $A$, including $\Ber A$, via the first
 $p+q$ traces $c_k=\Tr \wedge^k A$ ($k=1,2,\dots,p+q$), by a recursive procedure.
 However, equation \eqref{identity} for the differences  $\gamma_k=c_k-c_k^*$ gives much more.

Formula \eqref{identity} stands in the supercase instead of the equality \eqref{identityforminors}
holding in the ordinary case. This leads to highly non-trivial relations between exterior powers $\wedge^k V$
and $\Ber V\otimes \wedge^{p-q-k} V^*$ instead of the canonical isomorphism \eqref{canisomorphism}.

   Formula \eqref{identity} also gives a closed expression
  for $\Ber A$  in terms of traces.
  Indeed, it follows from
\eqref{superham1}--\eqref{denoteby} that
 \begin{equation}\label{beautiful}
    \Ber A=\Tr \wedge^{p-q} A-\gamma_{p-q}\,.
\end{equation}
(In the ordinary case $q=0$, $\gamma_{p-q}=0$ we arrive at
\eqref{determinantviatrace}.) Now, by considering
relation~\eqref{beautiful} and  identity~\eqref{identity} for
$k=p-q$ we arrive at the formula
         \begin{equation}\label{berintermsoftraces}
    \Ber A=\frac{\det\begin{pmatrix}
                  c_{p-q}(A) & \dots & c_p(A) \\
  \dots & \dots & \dots \\
  c_p(A) & \dots & c_{p+q}(A) \\
                \end{pmatrix}}{\det\begin{pmatrix}
  c_{p-q+2}(A) & \dots &  c_{p+1}(A) \\
  \dots & \dots & \dots \\
   c_{p+1}(A) & \dots &  c_{p+q}(A) \\
   \end{pmatrix}}\,.
\end{equation}
Here as before we set $c_k=0$ for $k<0$.

 For example, let $A$ be an even operator in a $p|1$-dimensional vector space.
 Then
\begin{equation*}\label{berintermsoftracesexample}
    \Ber A=\frac{\det\begin{pmatrix}
  c_{p-1}(A) & c_p(A) \\
  c_p(A) &  c_{p+1}(A) \\
                     \end{pmatrix}}{c_{p+1}(A)}=c_{p-1}(A)-\frac{c_p^2(A)}{c_{p+1}(A)}\,.
\end{equation*}

The rational expression in~\eqref{berintermsoftraces} is essentially
different from the original formula~\eqref{berezinian},  where the
numerator and denominator are not invariant functions of the matrix
$A$. Compared to it, the numerator and denominator of the fraction
in formula~\eqref{berintermsoftraces} are invariant polynomials.

  One can show that these invariant polynomials  are
  the traces of the representations  corresponding to certain Young diagrams.
 Namely, the numerator in  \eqref{berintermsoftraces}
 is equal to the trace of the action of the operator $A$ on an invariant subspace in
 the space of tensors $V^{\otimes N}$  corresponding to the rectangular
 Young diagram  $D_{p,q+1}$ with $p$ rows of length $q+1$.
Respectively, the denominator in \eqref{berintermsoftraces}
 is equal  to the trace of the action of the operator $A$ on an invariant subspace
 corresponding to the Young diagram  $D_{p+1,q}$
\footnote{If $A$ is an  ordinary $p\times p$ matrix,
 then by \eqref{determinantviatrace}, $\det A=c_p(A)$  simply equals the trace $\Tr\wedge^pA$  on the one-dimensional space of totally antisymmetric
 $p$-tensors,
 corresponding to the Young diagram $D_{p,1}$ with $p$ rows of length $1$.}.
This follows from the well-known  {\it Schur--Weyl formula}
\cite{weyl:classical} which can be generalised to the supercase (see
e.g. in \cite{kantor-trishin:cayley})

  Denote the invariant polynomials in the numerator and denominator of the fraction
in~\eqref{berintermsoftraces}  by $\Ber^+(A)$ and $\Ber^-(A)$,
respectively. What is the meaning of
   $\Ber^+(A), \Ber^-(A)$  in terms of the eigenvalues of the operator $A$?

  Compare \eqref{berintermsoftraces} with  expression
\eqref{ratiooftwopolynomials} for the characteristic function $R_A(z)$.
  Let $\{\l_1,\dots,\l_p\}$ and $\{\mu_1,\dots,\mu_q\}$
  be the eigenvalues of the even operator $A$ as above.
 Then consider the top coefficients $a_p,b_q$ of the polynomials $P(z), Q(z)$ in \eqref{ratiooftwopolynomials}.
 It follows that $a_p=\prod \lambda_i, b_q=\prod \mu_\a$, and $\Ber A={\prod \lambda_i\over \prod \mu_\a}$.
Hence $\Ber^+(A)=R\cdot a_p\,$, $\Ber^-(A)=R\cdot b_q\,$, with a
certain coefficient $R$. (Note that $a_p$ and $b_q$ are not
polynomials in the matrix entries of $A$.) One can explicitly find
$a_p\,$, $b_q$ by solving straightforwardly  a system of
simultaneous equations corresponding to the linear recurrence
relations \eqref{recurrentrealtionsright}. In particular, these
calculations give
         \begin{equation*}\label{resultant1}
    R=\det\begin{pmatrix}
            c_{p-q+1}(A) & \dots & c_p(A) \\
  \dots & \dots & \dots \\
  c_p(A) & \dots & c_{p+q-1}(A) \\
          \end{pmatrix}\,\,.
           \end{equation*}
   By considering \eqref{identity} one can come to an important observation that
            $R=\prod_{i,\a}(\lambda_i-\mu_a)$.
            Up to a sign it is just the classical Sylvester's resultant for the polynomials $P$ and $Q$
            standing at the top and bottom of the characteristic function
            $R_A(z)$ in \eqref{ratiooftwopolynomials}.
   Thus for the invariant polynomials  $\Ber^+(A), \Ber^-(A)$ we have:
   \begin{equation*}\label{polyninvariants}
    \Ber^+(A)=\prod_i \lambda_i \prod_{i,\a}(\l_i-\mu_\a)\,,\quad
\Ber^-(A)=\prod_\a \mu_\a\prod_{i,\a}(\l_i-\mu_\a)\,.
\end{equation*}

  The polynomials $\Ber^+(A), \Ber^-(A)$ appear in the analog
  of the Cayley--Hamilton theorem for the supercase.
  In particular, the  polynomial
       \begin{equation*}\label{CH}
    {\mathscr P}_A(z)={\Ber^+(A-z)\Ber^- (A-z)}\cdot\frac{1}{R}
\end{equation*}
is the minimal  annihilating polynomial for a generic even matrix $A$, and its coefficients are polynomial
invariants of $A$.

\end {document}